\documentclass{llncs}

\usepackage{makeidx}  
\usepackage{color}
\usepackage{amssymb}
\usepackage{amsmath}
\usepackage{amsfonts}
\usepackage{dsfont}
\usepackage{graphicx}

\newcommand{\XZ}{\mathds{Z}}
\newcommand{\XR}{\mathds{R}}

\newcommand{\XX}{\mathcal{X}}
\newcommand{\XC}{\mathcal{C}}
\newcommand{\XD}{\mathcal{D}}

\newcommand{\Cs}{C^\textrm{s}}
\newcommand{\Cb}{C^\textrm{b}}
\newcommand{\Xs}{X^\textrm{s}}
\newcommand{\Xb}{X^\textrm{b}}

\newcommand{\HPrep}{\textsc{HPrep}}
\newcommand{\latFit}{\mbox{\textsc{LatFit}}}
\newcommand{\SEQ}[1]{{\tiny #1}}
\newcommand{\Angstroms}{$\mathring{A}$}
\newcommand{\AVG}[1]{\langle #1 \rangle}
\newcommand{\neigh}{\mathcal{N}}
\newcommand{\relneigh}{\stackrel{\scriptscriptstyle{L}}{\sim}}

\title{Constraint-based Local Move Definitions for Lattice Protein Models
Including Side Chains}

\titlerunning{CP-based Local Moves}

\author{
	Martin Mann\inst{1} 
	\and Mohamed Abou Hamra\inst{2}
	\and Kathleen Steinh\"ofel\inst{3}
	\and Rolf Backofen\inst{1}
}
\authorrunning{Mann et al.}

\institute{
	University of Freiburg, Bioinformatics, 79110 Freiburg, Germany,\\
	\email{\{mmann,backofen\}@informatik.uni-freiburg.de},\\ 
	\and
	German University in Cairo, 5th Settlement New Cairo City, Cairo, Egypt,\\
	\email{mohamed.abdel-fattah@student.guc.edu.eg},\\ 
	\and
	King's College London, Department of Computer Science, London WC2R 2LS, UK,\\
	\email{kathleen.steinhofel@kcl.ac.uk},\\ 
}

\begin{document}

\maketitle


\begin{abstract}
The simulation of a protein's folding process is often done via stochastic local
search, which requires a procedure to apply structural changes onto a
given conformation. Here, we introduce a constraint-based approach to enumerate
lattice protein structures according to $k$-local moves in arbitrary lattices.
Our declarative description is much more flexible for extensions than
standard operational formulations. It enables a generic calculation of $k$-local
neighbors in backbone-only and side chain models. We exemplify the procedure
using a simple hierarchical folding scheme.
\end{abstract}

\section{Introduction}

The \emph{in silico} determination of a protein's functional fold is a well
established problem in bioinformatics. Since X-ray or NMR studies are still time
consuming and expensive, computational methods for \emph{ab initio} protein
structure prediction are needed. Despite research over the last decades, a
direct calculation of minimal energy structures in full atom resolution is
currently not feasible. Thus, heuristics and a wide variation of protein models
have been developed to identify fundamental principles guiding the process of
structure formation. A common abstraction of proteins are lattice protein
models~\cite{Backofen:06a,Kmiecik:07,Lau:89a,Mann_CPSPweb_2009}. Their
discretized structure space enables efficient folding
simulations~\cite{Ullah:09,Zhang:07} while maintaining good modelling
accuracy~\cite{Park:95}.

Folding simulations are often based on stochastic local searches, e.g. Monte
Carlo simulations~\cite{Ullah:09}. Different procedures, so called \emph{move
sets}, have been developed to calculate the structural changes along the
simulation, i.e. to enumerate the structural neighborhood of a certain
structure. A method often applied in literature are \emph{$k$-local
moves}~\cite{Sali:94b} that allow for structural changes within a successive
interval of fixed length~$k$. They are discussed in detail in Sec~\ref{sec-csp}.
Dotu and co-workers have used local moves for backbone-only HP~models
within a constraint-based large neighborhood search for optimal protein
structures~\cite{Dotu:08}. Lesh \emph{et al.} introduced \emph{pull
moves}~\cite{Lesh:03} that are widely used in recent
studies~\cite{Mann_LatPack_HFSP_08,Ullah:09}. \emph{Pivot moves} allow for the
rotation or reflection of subchains at an arbitrary Pivot position of the
structure~\cite{Madras:88}, while Zhang \emph{et al.} suggested a sequential
regrowth of structure fragments to enhance folding simulations~\cite{Zhang:07}.

All named move sets are currently restricted to backbone-only lattice
protein models, i.e. only the $C_\alpha$-trail of the protein is modeled. For
more realistic protein models incorporating side chains, often a combination of
different move sets is applied. Betancourt combined Pivot moves on the backbone
with a new FEM move set~\cite{Betancourt:05}, while Dima and Thirumalai have
used a combination of 2-local moves on the backbone with a simple relocation of
the side chain~\cite{Dima:02}. An exception is the advanced CABS model by Kolinski
and co-workers~\cite{Kmiecik:07}, which represents the side chain in higher
detail and requires more complex moves. 

Here, we introduce a generic and flexible approach to enable folding simulations
in backbone-only and side chain models using any $k$-local moves (i.e. any
interval length~$k$) in arbitrary lattices. The constraint programming (CP)
based formulation focuses on a description of the targeted structural neighbors
instead of an operational encoding of the moves possible. The introduced scheme
is therefore easy to extend with new directives or can be used for other
applications, e.g. fragment re-localization~\cite{Zhang:07}, as discussed later.
Beneath applications in studies of the whole energy
landscape~\cite{Mann_ELL_BIRD07}, the approach is well placed to be applied
within a local search following the framework of Pesant and
Gendreau~\cite{Pesant:99}. We apply our move set for side chain models within
a simple folding simulation procedure in the style of~\cite{Ullah:09} and evaluate
the results with known protein structures.

\section{Preliminaries}
\label{sec-prelim}

\begin{figure}[t]
\vspace{-1em}
\begin{center}
	a)
	\begin{minipage}[t]{0.3\textwidth}
		\includegraphics[width=\textwidth]{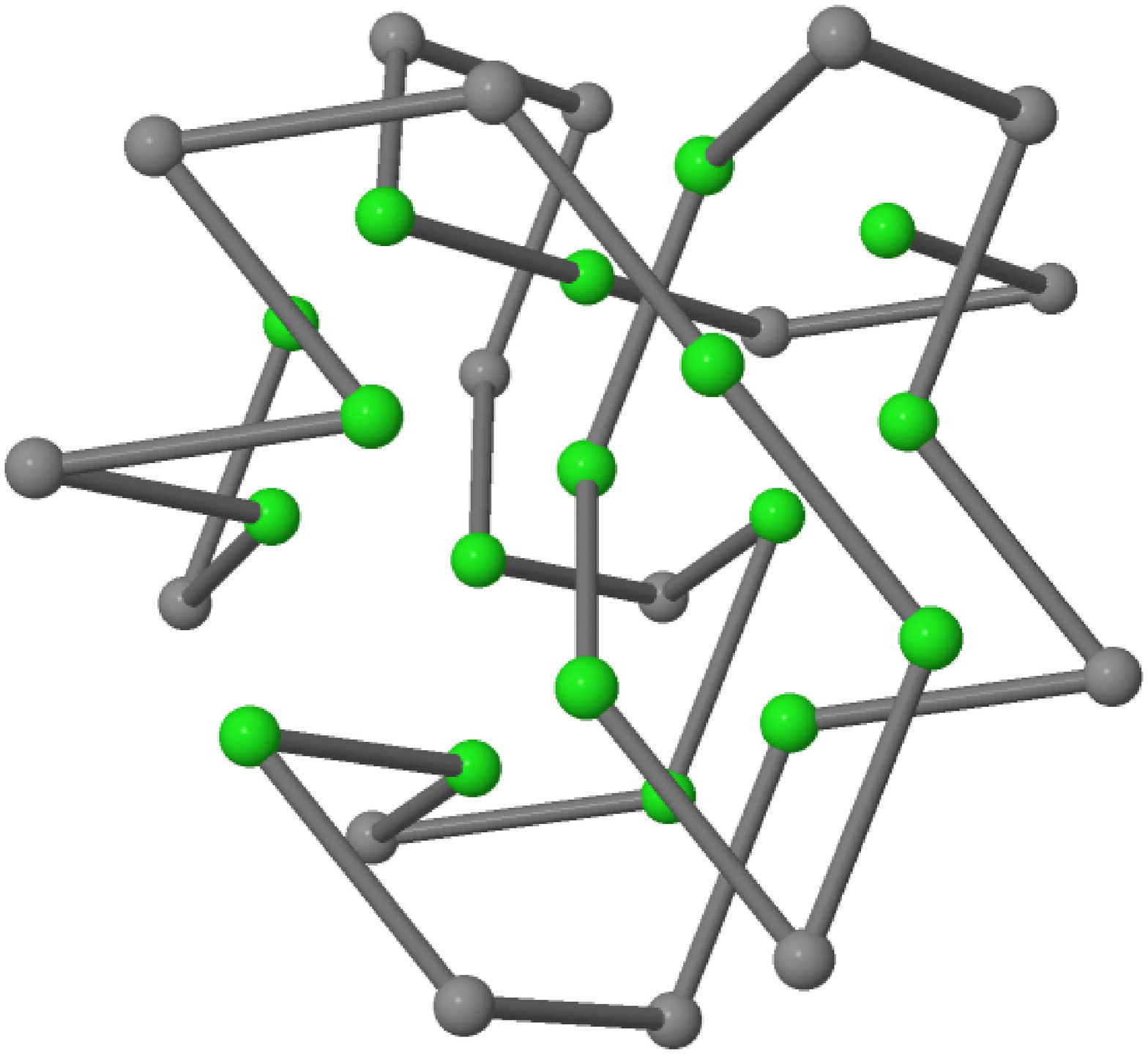}
    \end{minipage}
	\hspace{2em}
	b)
	\begin{minipage}[t]{0.3\textwidth}
		\includegraphics[width=\textwidth]{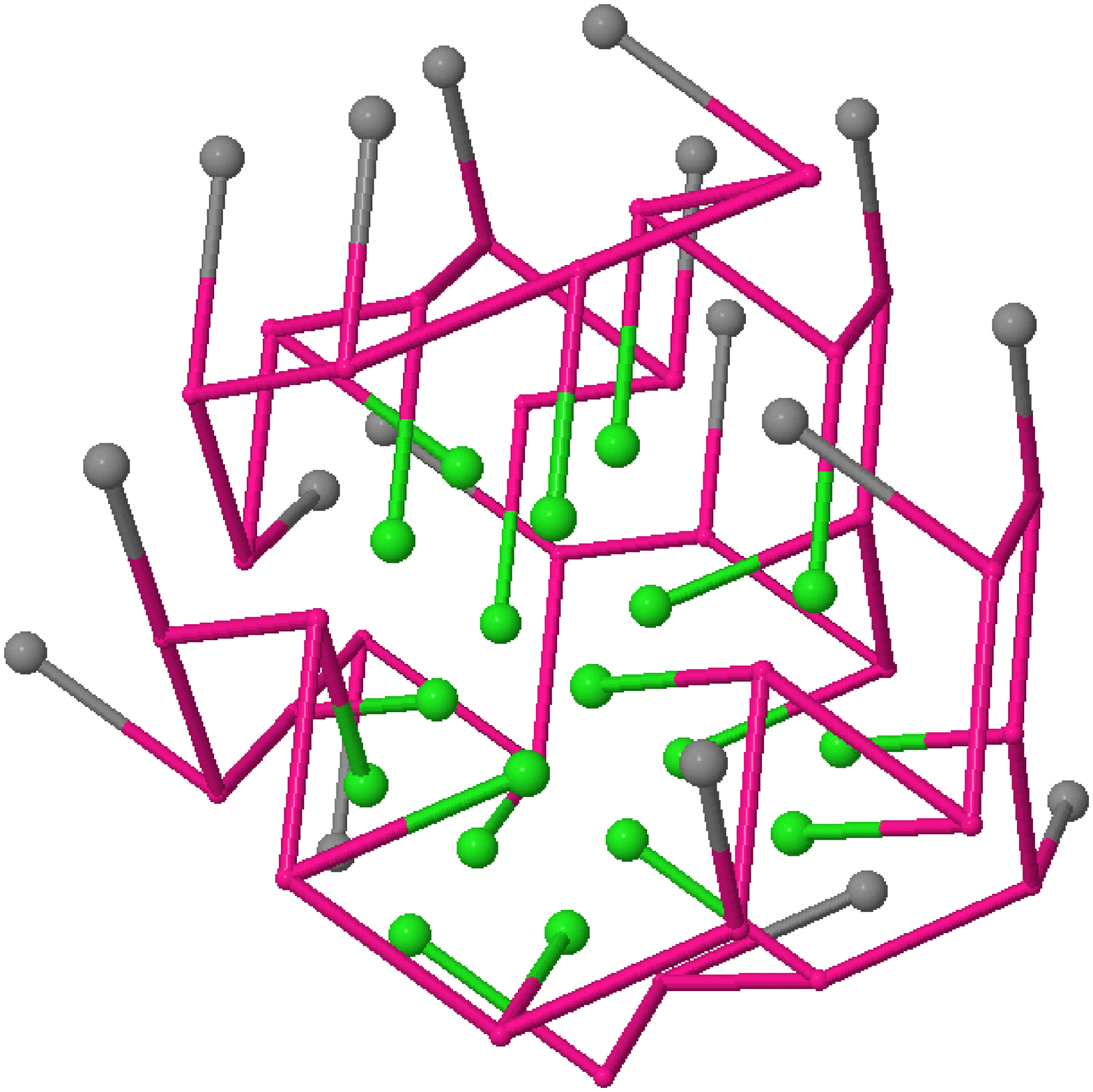}
    \end{minipage}
	\vspace{-0.5em}
	\caption{HP-optimal structures of
	{\small\texttt{HPPHHPPPHPHHPHHPPHPHPPHHHPHHPPHPHPH}} in the
	face-centered-cubic lattice. (a) backbone-only model with energy~-50,
	(b) side chain model with energy~-55. Colors: green - H~monomers, 
	gray - P~monomers, red - backbone in side chain models.
	Calculation and visualization are done using the
	CPSP-package~\cite{Mann:08a}.}
	\label{fcc-struct}
\end{center}
\vspace{-2em}
\end{figure}

Given a lattice~$L\subseteq \XZ^3$ and an according neighborhood
relation~$\relneigh$ between coordinates of~$L$. A \emph{backbone-only} lattice
protein of length~$n$ is described by~$(S,C)$ where~$S\in \Sigma^n$ denotes the
sequence over some alphabet~$\Sigma$ (e.g. the 20~proteinogen amino acids)
and~$C\in L^n$ the lattice nodes occupied. A valid lattice protein structure
satisfies connectivity of successive monomers $\forall_{1\leq i<n}: C_i
\relneigh C_{i+1}$ and their self-avoidingness $\forall_{1\leq i<j\leq n}:C_i
\neq C_j$. A \emph{side chain} lattice protein is defined by~$(S,\Cb,\Cs)$, i.e.
a sequence~$S\in \Sigma^n$, the backbone positions~$\Cb\in L^n$ and the side
chain positions~$\Cs\in L^n$. The side chain position~$\Cs$ represents the
centroid of the amino acid's side chain atoms. A valid lattice protein structure
including side chains satisfies connectivity of successive backbone monomers
$\forall_{1\leq i<n}: \Cb_i \relneigh \Cb_{i+1}$, the connection of backbone and
side chain for each amino acid $\forall_{1\leq i\leq n}: \Cb_i \relneigh \Cs_i$,
and the selfavoidingness of all monomers $\forall_{1\leq i<j\leq n}: \Cb_i \neq
\Cb_j \wedge \Cs_i \neq \Cs_j \wedge \Cb_i \neq \Cs_j \wedge \Cb_i \neq \Cs_i$.
We consider the contact based energy functions~$E^\textrm{b}(S,C) = \sum_{1\leq
i<j\leq n}^{(C_i \relneigh\; C_j)} e(S_i,S_j)$ for backbone-only and
$E^\textrm{s}(S,\Cb,\Cs) = \sum_{1\leq i<j\leq n}^{(\Cs_i \relneigh\; \Cs_j)}
e(S_i,S_j)$ for side chain lattice proteins for a given energy contribution
function~$e : \Sigma \times \Sigma \rightarrow \XR$. Note, the energy function
for side chain proteins considers (as in~\cite{Bromberg:94}) the contacts
between side chain positions only! $e^{20}$ denotes an empirical 20 amino acid
contact potential as described in~\cite{Berrera:03,DalPalu:04b}. $e^{HP}$
represents the energy contribution function of the Hydrophobic-Polar
(HP)~model~\cite{Lau:89a}, i.e. it returns~$-1$ if both amino acids are
hydrophobic, $0$~otherwise. Our hydrophobic/polar (H/P) assignment
follows~\cite{Ullah:09}. An \emph{optimal structure} minimizes the energy
function. In the following, we denote a structure \emph{HP-optimal} if it
minimizes the energy function based on~$e^{HP}$. Figure~\ref{fcc-struct}
exemplifies HP-optimal structures for both lattice protein models. In the
following, we assume a scaled lattice such that neighbored positions in the
lattice have a distance of~3.8\Angstroms, the average $C_\alpha$-atom distance
in proteins.

\section{Constraint-based Local Move Set Definition}
\label{sec-csp}

To enable folding simulations we need a definition of structural changes that
encodes the structural neighborhood of a given lattice protein structure.
Here, we follow the idea of $k$-local moves, that confine the
difference between the initial and the neighbored structure to a consecutive
interval of maximal length of~$k$. Therefore, we define the
$k$-neighborhood~$\neigh_k(C)$ of a given structure~$C$ as:
\begin{equation}
	\neigh_k(C) = 
	\left\{
		\text{valid structures }C' \;|\;
		\exists_{1\leq s \leq n} : 
		\forall_{j \not\in [s,\ldots,(s+k-1)]} : C_j = C'_j
	\right\}
	\label{eq-N_k}
\end{equation}

In order to enumerate all valid structural neighbors~$C' \in \neigh_k(C)$ of a
given lattice protein~$C$, we have to enumerate the neighbors for all possible
interval lengths~$1\leq k'\leq k$ and interval starts~$1\leq s \leq(n-k'+1)$.
Since we want to calculate each neighboring structure only once, we have to
enhance the $k$-local move definition to \emph{strict $k$-local moves}. Here, we
enforce in addition that both ends ($C'_s$ and $C'_{s+k-1}$) of the successive
interval of length~$k$ are changed, i.e. a strict $k$-local move does not cover
a $k'$-local move with $k'<k$, as a normal $k$-local move in accordance with
Eq.~\ref{eq-N_k} does. This ensures a unique enumeration of structural
neighbors for an increasing~$k'$.

In the following, we will introduce the Constraint Satisfaction Problems (CSP)
that describe all valid structural neighbors~$C' \in \neigh_k(C)$ of a given
lattice protein~$C$ according to strict $k$-local moves in a lattice~$L$. A CSP
is given by $(\XX,\XD,\XC)$, where we denote the set of variables~$\XX$, their
domains~$\XD$, and a set of constraints~$\XC$. A solution of a CSP is an
assignment $a_i\in\XD(X_i)$ for each variable that satisfies all constraints
in~$\XC$. To simplify the presentation, we utilize a binary neighboring
constraint $\operatorname{neigh}(X,Y)$ that ensures $\forall_{d_x\in \XD(X)} :
\exists_{d_y\in \XD(Y)} : (d_x \relneigh d_y)$ and vice versa. Furthermore, we
use the global $\operatorname{all-different}$ constraint by
R\'egin~\cite{Regin:94} to enforce pairwise differences within a set of
variables.

\subsection{CSP for Backbone-only Models}

Given a valid backbone-only lattice protein structure~$C$ of length~$n$, a move
interval length~$k\leq n$, and the start of the interval~$1\leq s\leq(n-k+1)$.
We define $k$~variables~$X_i$, one for each position of the interval, with
$\XD(X_i) = L \setminus \{C_1,\ldots,C_{s-1},C_{s+k},\ldots,C_n\}$. These
variables have to form a valid structure, therefore we post
$\operatorname{all-different}(X_1,\ldots,X_k)$ and $\forall_{1\leq i<k}:
\operatorname{neigh}(X_i,X_{i+1})$. Since we describe a substructure, it has to
be connected to the interval borders: if $s>1 :
\operatorname{neigh}(X_1,C_{s-1})$ and if $(s+k-1)<n
:\operatorname{neigh}(X_k,C_{s+k})$. Finally, we enforce that both ends of the
interval are different from the old placement, i.e. $X_1 \neq C_i$ and $X_k \neq
C_{i+k-1}$, to enumerate strict $k$-local move neighbors only.

The presented CSP is similar to the work of Dotu \emph{et al.}~\cite{Dotu:08},
but in contrast ensures the uniqueness of each move. Thus, each neighbored
structure is available only via a single interval. This is of high importance to
enable a non-redundant enumeration of a structure's neighborhood in the fold
space to access its energy landscape~\cite{Mann_ELL_BIRD07}.

\subsection{CSP for Models Including Side Chains}

Given a valid side chain lattice protein structure~$(\Cb,\Cs)$ of length~$n$, a
move interval length~$k\leq n$, and the start of the interval~$1\leq
i\leq(n-k+1)$. We define $k$~variables~$\Xb_i$ and~$\Xs_i$, two for each
position of the interval, with $\XD(\Xb_i) = \XD(\Xs_i) =
L\setminus\{\Cb_1,\ldots,\Cb_{s-1},\Cb_{s+k},\ldots,\Cb_n,\Cs_1,\ldots,\Cs_{s-1},\Cs_{s+k},\ldots,\Cs_n\}$.
To ensure a valid structure, we enforce
$\operatorname{all-different}(\Xb_1,\ldots,\Xb_k,\Xs_1,\ldots,\Xs_k)$,
$\forall_{1\leq i<k}: \operatorname{neigh}(\Xb_i,\Xb_{i+1})$, and
$\forall_{1\leq i\leq k}: \operatorname{neigh}(\Xb_i,\Xs_i)$. Since we describe
a substructure, it has to be connected to the interval borders: if $s>1 :
\operatorname{neigh}(\Xb_1,\Cb_{s-1})$ and if $(s+k-1)<n
:\operatorname{neigh}(\Xb_k,\Cb_{s+k})$. Finally, we warrant the strictness of
the $k$-local moves and enforce that both ends of the interval differ from the
old backbone or side chain placement, i.e. $(\Xb_1 \neq \Cb_i \vee \Xs_1 \neq
\Cs_i)$ and $(\Xb_k \neq \Cb_{i+k-1} \vee \Xs_k \neq \Cs_{i+k-1})$.

\section{Application}
\label{sec-appl}

In the following, we applied the introduced move set to folding simulations of
side chain lattice protein models in the 3D~face-centered-cubic (FCC) lattice.
In the FCC lattice, two lattice points~$l_1$ and~$l_2$ are neighbored, if and
only if\\ $(l_1-l_2)\in\left\{\pm(1,1,0),\pm(1,0,1),\pm(0,1,1),\pm(1,-1,0),\pm(1,0,-1),\pm(0,1,-1)\right\}$.\\
Thus, each point of the FCC lattice has 12~neighbored positions. $k$-local moves
are known to be non-ergodic for backbone-only
models~\cite{Madras:87} depending on~$k$, the used lattice, and
the protein length. We expect the same for models including side chains, but
using the FCC and an intermediate~$k$ should shift the problem to long chain
lengths. Thus, we apply 3-local moves, i.e. with a maximal interval length~$k=3$
such that up to 6~monomers are moved (2~per amino acid). The implementation is
based on Gecode~\cite{GecodeWebsite}. To evaluate the structural difference
between two structures~$(\Cb,\Cs)$ and~$(\hat{\Cb},\hat{\Cs})$ we calculate the
distance and coordinate root mean square deviation (dRMSD and cRMSD) as given by
Eq.~\ref{dRMSD} and~\ref{cRMSD}, respectively. The needed superpositioning
utilizes Kabsch's algorithm~\cite{Kabsch:76}. We apply the contact based energy
function~$E^\textrm{s}$ that evaluates (only) side chain monomer contacts using
the $e^{20}$~contact energy potentials from Sec.~\ref{sec-prelim} similar to the
backbone-only studies in~\cite{Berrera:03,Ullah:09}. In the following, we
use~$C$ as an abbreviation for~$(\Cb,\Cs)$. { \footnotesize
\begin{eqnarray}
 	\text{dRMSD} &:&
	\sqrt{
	 \frac{
	  \begin{matrix}
		\sum_{i<j}
		(|\Cb_i-\Cb_j|-|\hat{\Cb}_i-\hat{\Cb}_j|)^2 +
		(|\Cs_i-\Cs_j|-|\hat{\Cs}_i-\hat{\Cs}_j|)^2
	  \\
		 + \sum_{i}
		(|\Cb_i-\Cs_i|-|\hat{\Cb}_i-\hat{\Cs}_i|)^2
	  \end{matrix}
	 }{n^2}
	} 
	\label{dRMSD} 
	\\
 	\text{cRMSD} &:&
	\sqrt{
	 \frac{
		\sum_{i}
		(|\Cb_i-\hat{\Cb}_i|)^2 +
		(|\Cs_i-\hat{\Cs}_i|)^2
	 }{2\cdot n}
	} 
	\label{cRMSD} 
\end{eqnarray}
}
We derived a protein data from the Pisces web server~\cite{Wang:03} on June~23rd
2009. Only complete X-ray structures of 2.0\Angstroms~resolution or better with
an R-value of~0.3 that contain side-chain data were considered. We used a 30\%
sequence identity cut-off. Since we are applying a simple contact-based energy
function we filtered for short globular shaped proteins. Table~\ref{tab-data}
summarizes the used sequences and their corresponding Protein Data Bank (PDB)
identifiers etc.


\begin{table}[tb]
\begin{center}
\small 
\begin{tabular}{c|ll}
	{\small PDB ID - chain} 
	 && Sequences (original and HP transform)
	\\
	\hline
    1BAZ-A
     && \SEQ{SKMPQVNLRWPREVLDLVRKVAEENGRSVNSEIYQRVMESFKKEGRIGA} \\
     && \SEQ{PPHPPHPHPHPPPHHPHHPPHPPPPPPPHPPPHHPPHHPPHPPPPPHPP} \\
    1J8E-A 
     && \SEQ{GSHSCSSTQFKCNSGRCIPEHWTCDGDNDCGDYSDETHANCTNQ} \\
     && \SEQ{PPPPHPPPPHPHPPPPHHPPPHPHPPPPPHPPHPPPPPPPHPPP} \\
    1RH6-A 
     && \SEQ{MYLTLQEWNARQRRPRSLETVRRWVRESRIFPPPVKDGREYLFHESAVKVDLNRP} \\
     && \SEQ{HHHPHPPHPPPPPPPPPHPPHPPHHPPPPHHPPPHPPPPPHHHPPPPHPHPHPPP} \\
    1Z0J-B 
     && \SEQ{IEEELLLQQIDNIKAYIFDAKQCGRLDEVEVLTENLRELKHTLAKQKGGTD} \\
     && \SEQ{HPPPHHHPPHPPHPPHHHPPPPHPPHPPHPHHPPPHPPHPPPHPPPPPPPP} \\
    2DS5-A 
     && \SEQ{GKLLYCSFCGKSQHEVRKLIAGPSVYICDECVDLCNDIIREEI} \\
     && \SEQ{PPHHHHPHHPPPPPPHPPHHPPPPHHHHPPHHPHHPPHHPPPH} \\
    2EQ7-C 
     && \SEQ{LAMPAAERLMQEKGVSPAEVQGTGLGGRILKEDVMRH} \\
     && \SEQ{HPHPPPPPHHPPPPHPPPPHPPPPHPPPHHPPPHHPP} \\
    2HBA-A 
     && \SEQ{MKVIFLKDVKGMGKKGEIKNVADGYANNFLFKQGLAIEATPANLKALEAQKQ} \\
     && \SEQ{HPHHHHPPHPPHPPPPPHPPHPPPHPPPHHHPPPHPHPPPPPPHPPHPPPPP} \\
\end{tabular}
\\
\vspace{1em}
\begin{tabular}{l|c||c|c||c|c||c|c||}
	PDB ID 
	 &
	 & \multicolumn{2}{c||}{$C_{fit}$ to $C_{PDB}$} 
	 & \multicolumn{2}{c||}{}
	 & \multicolumn{2}{c||}{$g(C_{fit})$ to $C_{fit}$} 
	\\
	- chain & $n$ 
	 & dRMSD & cRMSD 
	 & $E(C_{fit})$ & $E(g(C_{fit}))$ 
	 & dRMSD & cRMSD 
	\\
	\hline
    1BAZ-A & 49
     & 0.886 \Angstroms & 1.725 \Angstroms
         & -3.73 & -31.51
	 & 4.050 \Angstroms & 6.565 \Angstroms
    \\
    1J8E-A & 44
     & 0.928 \Angstroms & 1.939 \Angstroms
         & -3.54 & -30.76
	 & 3.865 \Angstroms & 6.857 \Angstroms
    \\
    1RH6-A & 55
     & 0.921 \Angstroms & 1.791 \Angstroms
         & 1.33 & -38.17
	 & 4.192 \Angstroms & 8.243 \Angstroms
    \\
    1Z0J-B & 51
     & 0.917 \Angstroms & 2.095 \Angstroms
         & 2.05 & -35.95
	 & 3.185 \Angstroms & 6.640 \Angstroms
    \\
    2DS5-A & 43
     & 0.901 \Angstroms & 1.750 \Angstroms
         & -4.35 & -34.36
	 & 4.658 \Angstroms & 7.755 \Angstroms
    \\
    2EQ7-C & 37
     & 0.905 \Angstroms & 1.813 \Angstroms
         & -3.07 & -20.58
	 & 2.328 \Angstroms & 4.751 \Angstroms
    \\
    2HBA-A & 52
     & 0.890 \Angstroms & 1.780 \Angstroms
         & -3.04 & -30.62
	 & 3.224 \Angstroms & 6.015 \Angstroms
    \\
\end{tabular}
\vspace{1em}
\caption{Used sequences, their HP transforms, length~$n$, the quality of the
fitted lattice protein model, and the according energies.}
\vspace{-3em}
\label{tab-data}
\end{center}
\end{table}

For each full atom PDB structure~$C_{PDB}$, we derived a lattice protein
structure~$C_{fit}$ that minimizes the dRMSD to~$C_{PDB}$. This was done using
\latFit{} from the \textsc{LatPack}-tools package v1.7.0\footnote{Freely
available at
http://www.bioinf.uni-freiburg.de/Software/LatPack/}~\cite{Mann_LatPack_HFSP_08}.
Table~\ref{tab-data} summarizes the resulting dRMSD and cRMSD values.

Since the applied energy function is still a rough abstraction of the forces
that guide the real folding process into~$C_{PDB}$, no energy minimizing folding
strategy will find the fitted lattice protein structure~$C_{fit}$. Thus, we
map~$C_{fit}$ to the according local minimum in the energy landscape. The
mapping is done via a steepest decent or \emph{gradient walk}. Starting from a
given structure, at each step the neighbored structure with lowest energy is
chosen for the next step until no such neighbor exists. Therefore, a gradient
walk ends in a local minimum of the energy landscape, which we denote~$g(C)$ for
a given start structure~$C$.

The~$g(C_{fit})$ structures represent our ``true'' model to benchmark the
following folding scheme. The energies of~$C_{fit}$ and~$g(C_{fit})$ and their
structural differences to each other and to~$C_{PDB}$ are given in
Table~\ref{tab-data}.

The folding simulation procedure applied follows the idea of~\cite{Ullah:09}.
For each amino acid sequence~$S$, we derive an according HP-sequence~$S_{HP}$
using the translation table used in~\cite{Ullah:09}. The derived~$S_{HP}$ are
given in Table~\ref{tab-data}. Following the observation of the hydrophobic
collapse~\cite{Agashe:95}, we calculated HP-optimal structure representatives
utilizing the CPSP-approach~\cite{Backofen:06a,Mann:08a,Mann_CPSPweb_2009} and
its latest extension \HPrep{}~\cite{Mann:HPrep:09}. The resulting HP-optimal
structures are named~$C_{HP}$. For each~$C_{HP}$ we run gradient walks and
evaluated the resulting local minima found. The corresponding energies are
listed in Table~\ref{tab-E-RMSD}. Furthermore, we performed a structural
comparison of the resulting $g(C_{HP})$~structures to our ``true''
models~$g(C_{fit})$ from the fitting. The RMSD values are given in
Table~\ref{tab-E-RMSD}.

In addition, we executed for each~$C_{HP}$ \emph{random descending walks} in
order to sample the local minima of the energy landscape accessible from the
collapsed starting structures. Here, at each step a random neighbor with lower
energy is selected following a uniform distribution until no such neighbor
exists. The lowest reached local minimum of all random descending walks starting
at~$C$ is denoted by~$r(C)$. Energy and structural differences are
given in Table~\ref{tab-E-RMSD}.

\begin{table}[tb]
\begin{center}
\small 
\begin{tabular}{l||c||c|c|}
	PDB ID  
	& \multicolumn{1}{c||}{ average values }
	& \multicolumn{2}{c|}{ minimal values }
	\\
	- chain
	 & $\AVG{E(C_{HP})}$
	 & $\min{E(g(C_{HP}))}$
	 & $\min{E(r(C_{HP}))}$
	\\
	\hline
    1BAZ-A
	 & -10.67
     & -33.07 & -34.60
    \\
    1J8E-A 
	 & -12.45
     & -29.33 & -32.35
    \\
    1RH6-A 
	 & -13.09
     & -35.12 & -37.59
    \\
    1Z0J-B 
	 & -13.42
     & -34.71 & -37.69
    \\
    2DS5-A 
	 & -6.97
     & -31.00 & -32.53
    \\
    2EQ7-C 
	 & -6.55
     & -21.64 & -25.10
    \\
    2HBA-A 
	 & -11.07
     & -30.91 & -35.56
    \\
\end{tabular}
\\
\vspace{1em}
\begin{tabular}{l||c|c||c|c||}
	PDB ID
	& \multicolumn{2}{c||}{ $g(C_{HP})$ vs. $g(C_{fit})$ }
	& \multicolumn{2}{c||}{ $r(C_{HP})$ vs. $g(C_{fit})$ }
	\\
	- chain
	 & dRMSD & cRMSD
	 & dRMSD & cRMSD
	\\
	\hline
    1BAZ-A
	 & 4.736 \Angstroms & 8.797 \Angstroms
	 & 4.762 \Angstroms & 9.360 \Angstroms
    \\
    1J8E-A
	 & 3.384 \Angstroms & 7.508 \Angstroms
	 & 3.196 \Angstroms & 7.052 \Angstroms
    \\
    1RH6-A 
	 & 4.190 \Angstroms & 9.645 \Angstroms
	 & 4.242 \Angstroms & 10.156 \Angstroms
    \\
    1Z0J-B
	 & 5.609 \Angstroms & 10.166 \Angstroms
	 & 6.232 \Angstroms & 11.438 \Angstroms
    \\
    2DS5-A 
	 & 3.588 \Angstroms & 8.679 \Angstroms
	 & 3.425 \Angstroms & 7.639 \Angstroms
    \\
    2EQ7-C 
	 & 3.427 \Angstroms & 7.247 \Angstroms
	 & 4.177 \Angstroms & 8.401 \Angstroms
    \\
    2HBA-A 
	 & 3.832 \Angstroms & 8.848 \Angstroms
	 & 4.194 \Angstroms & 9.075 \Angstroms
    \\
\end{tabular}
\vspace{1em}
\caption{Resulting energies and a structural comparison of the folding results.}
\vspace{-3em}
\label{tab-E-RMSD}
\end{center}
\end{table}

\section{Discussion}

The gradient walks using 3-local moves starting from the fitted
structures~$C_{fit}$ revealed that the currently applied contact based energy
function using the energy potentials~$e^{20}$, originally derived for
backbone-only models~\cite{Berrera:03}, does not reflect the real forces present
for models including side chains. This can be observed when comparing the
energies~$E(C_{fit})$ to~$E(g(C_{fit}))$ (see Table~\ref{tab-data}). An energy
function that results in a smaller difference would be preferable, i.e. it would
be a better model for the real forces guiding the folding process to~$C_{PDB}$.
In addition we could show, that the derived structures from our simple
energy-optimizing folding simulation procedure are still quite dissimilar to the
energy-optimized lattice fits of the real structures (see
Table~\ref{tab-E-RMSD}). We assume this mainly results from the simple energy
function as well.

To improve the results, we plan to apply more advanced energy functions, e.g.
following~\cite{Kmiecik:07}. Most important: the energy function has to consider
the backbone positioning as well, which is not done by the contact-based energy
functions from Sec.~\ref{sec-prelim}. Additionally, we want to apply
distance based energy potentials that allow for a more realistic energy
evaluation. Another direction of ongoing research is to further constrain the
allowed structures. Here, we will directly benefit from the CP-based formulation
of $k$-local moves. Since we are formulating a CSP on valid structural
neighbors, it is quite easy to post additional structural constraints. For
instance, we can enforce a restriction on the allowed relative torsion angles
along the protein chain (as e.g. done in~\cite{DalPalu:04}), that follows the
observation of a limited degree of freedom in nature.

\section{Conclusions and Summary}

We introduced a CP-based approach to enumerate $k$-local neighbors of a lattice
protein structure in backbone-only and side chain lattice protein models. The
generic approach can be applied for any local move length~$k$ within arbitrary
lattices. Thus, it enables a fast prototyping of new folding simulation schemes
or can be easily extended with additional constraints, e.g. restricted torsion
angles. The CSP formulation enables the enumeration of the whole $k$-local move
neighborhood~$\neigh_k(C)$ of a given structure~$C$ or the calculation of a
random neighboring structure~$C_r\in\neigh_k(C)$ when applying a randomized
search as possible in Gecode~\cite{GecodeWebsite}. The application of symmetry
breaking search~\cite{Backofen:02} can be used to avoid the enumeration of
symmetric structures, increasing the efficiency of folding
simulations~\cite{Gan:08}. We plan the incorporation of the $k$-local move
neighbor enumeration into our C++ energy landscape library
(ELL)~\cite{Mann_ELL_BIRD07}. This will open an easy interface for folding
simulations in arbitrary lattices utilizing any energy function of interest.
Furthermore, this will enable full kinetics studies based on the energy
landscape topology.

We will utilize the flexibility of the CP-based approach to incorporate
additional structural constraints into the neighborhood generation.
Following~\cite{DalPalu:04,DalPalu:04b}, it is beneficial to restrict torsion
angles along the backbone or to exploit secondary structure information.

Another advantage of the CP-based approach is its extensibility to constraint
optimization problems (COP). Currently, we plan to incorporate the energy
function as the objective into the CSP, as e.g. done
in~\cite{Cipriano:08,Dotu:08}. Thus, by solving a COP while optimizing the
energy function, we can directly calculate the lowest energy neighbor of a
structure following the framework of Pesant and
Gendreau~\cite{Pesant:99}, which is needed e.g. for a gradient walk in the
energy landscape as done in Sec.~\ref{sec-appl}. Furthermore, this would enable
an extension of the work of Zhang \emph{et al.}~\cite{Zhang:07}. They showed
(for backbone-only models) that the performance of Monte Carlo folding
simulations can be significantly increased using a greedy sequential regrowth of
subchains. Thus, we plan to directly apply the sketched COP to calculate the
optimal fragments for lattice proteins including side chains. Finally, the
presented CP-based move set formulation can be easily extended to any other
local move definition of interest.


\begin{thebibliography}{10}

\bibitem{Agashe:95}
Vishwas~R. Agashe, M.~C.~R. Shastry, and Jayant~B. Udgaonkar.
\newblock Initial hydrophobic collapse in the folding of barstar.
\newblock {\em Nature}, 377:754--757, 1995.

\bibitem{Backofen:02}
Rolf Backofen and Sebastian Will.
\newblock Excluding symmetries in constraint-based search.
\newblock {\em Constraints}, 7(3):333--349, 2002.

\bibitem{Backofen:06a}
Rolf Backofen and Sebastian Will.
\newblock A constraint-based approach to fast and exact structure prediction in
  three-dimensional protein models.
\newblock {\em J Constraints}, 11(1):5--30, Jan 2006.

\bibitem{Berrera:03}
Marco Berrera, Henriette Molinari, and Federico Fogolari.
\newblock Amino acid empirical contact energy definitions for fold recognition
  in the space of contact maps.
\newblock {\em BMC Bioinformatics}, 4:8, 2003.

\bibitem{Betancourt:05}
M.~R. Betancourt.
\newblock Efficient monte carlo trial moves for polypeptide simulations.
\newblock {\em J Chem Phys}, 123(17), 2005.

\bibitem{Cipriano:08}
Raffele Cipriano, Alessandro~Dal Pal\`u, and Agostino Dovier.
\newblock A hybrid approach mixing local search and constraint programming
  applied to the protein structure prediction problem.
\newblock In {\em Proc of WCB'08}, page~?, 2008.

\bibitem{Bromberg:94}
S.~Bromberg K.~A. Dill.
\newblock Side-chain entropy and packing in proteins.
\newblock {\em Protein Sci}, 3(7):997--1009, 1994.

\bibitem{Dima:02}
R.I. Dima and D.~Thirumalai.
\newblock Exploring protein aggregation and self-propagation using lattice
  models: Phase diagram and kinetic.
\newblock {\em Prot. Sci.}, 11(5):1036--–1049, 2002.

\bibitem{Dotu:08}
Ivan Dotu, Manuel Cebri\'an, Pascal~Van Hentenryck, and Peter Clote.
\newblock Protein structure prediction with large neighborhood constraint
  programming search.
\newblock In {\em Proc of CP'08}, volume 5202 of {\em LNCS}, pages 82--96.
  Springer, 2008.

\bibitem{Gan:08}
Xiangchao Gan, Leonidas Kapsokalivas, Andreas~A. Albrecht, and Kathleen
  Steinh\"ofel.
\newblock A symmetry-free subspace for ab initio protein folding simulations.
\newblock In {\em Proc. of BIRD'08}, volume~13 of {\em CCIS}, pages 128--139.
  Springer, 2008.

\bibitem{GecodeWebsite}
{Gecode}: Generic constraint development environment, 2007.
\newblock Available as an open-source library from \texttt{www.gecode.org}.

\bibitem{Kabsch:76}
W.~Kabsch.
\newblock A solution for the best rotation to relate two sets of vectors.
\newblock {\em Acta Crystallographica}, A32:922--923, 1976.

\bibitem{Kmiecik:07}
Sebastian Kmiecik and Andrzej Kolinski.
\newblock Characterization of protein-folding pathways by reduced-space
  modeling.
\newblock {\em PNAS}, 104(30):12330--12335, 2007.

\bibitem{Lau:89a}
Kit Lau and Ken Dill.
\newblock A lattice statistical mechanics model of the conformational and
  sequence spaces of proteins.
\newblock {\em Macromolecules}, 22(10):3986--3997, 1989.

\bibitem{Lesh:03}
Neal Lesh, Michael Mitzenmacher, and Sue Whitesides.
\newblock A complete and effective move set for simplified protein folding.
\newblock In {\em Proc of RECOMB'03}, pages 188--195. ACM, 2003.

\bibitem{Madras:87}
Neal Madras and Alan~D. Sokal.
\newblock Nonergodicity of local, length-conserving {Monte} {Carlo} algorithms
  for the self-avoiding walk.
\newblock {\em Journal of Statistical Physics}, 47(3-4):573--595, 1987.

\bibitem{Madras:88}
Neal Madras and Alan~D. Sokal.
\newblock The pivot algorithm: A highly efficient {Monte Carlo} method for the
  self-avoiding walk.
\newblock {\em J Stat Phys}, 50(1-2):109--186, 1988.

\bibitem{Mann:HPrep:09}
Martin Mann, Rolf Backofen, and Sebastian Will.
\newblock Equivalence classes of optimal structures in {HP}~protein models
  including side chains.
\newblock In {\em Proc of WCB'09}, 2009.

\bibitem{Mann_LatPack_HFSP_08}
Martin Mann, Daniel Maticzka, Rhodri Saunders, and Rolf Backofen.
\newblock Classifying protein-like sequences in arbitrary lattice protein
  models using {LatPack}.
\newblock {\em HFSP Journal}, 2(6):396, 2008.

\bibitem{Mann_CPSPweb_2009}
Martin Mann, Cameron Smith, Mohamad Rabbath, Marlien Edwards, Sebastian Will,
  and Rolf Backofen.
\newblock {CPSP-web-tool} : a server for {3D} lattice protein studies.
\newblock {\em Bioinformatics}, 25(5):676--677, 2009.

\bibitem{Mann_ELL_BIRD07}
Martin Mann, Sebastian Will, and Rolf Backofen.
\newblock The energy landscape library - a platform for generic algorithms.
\newblock In {\em Proc of BIRD'07}, volume 217, pages 83--86. OCG, 2007.

\bibitem{Mann:08a}
Martin Mann, Sebastian Will, and Rolf Backofen.
\newblock {CPSP}-tools - exact and complete algorithms for high-throughput {3D}
  lattice protein studies.
\newblock {\em BMC Bioinformatics}, 9:230, 2008.

\bibitem{DalPalu:04b}
Alessandro~Dal Palu, Agostino Dovier, and Federico Fogolari.
\newblock Constraint {Logic} {Programming} approach to protein structure
  prediction.
\newblock {\em BMC Bioinformatics}, 5(1):186, 2004.

\bibitem{DalPalu:04}
Alessandro~Dal Pal\`u, Sebastian Will, Rolf Backofen, and Agostino Dovier.
\newblock Constraint based protein structure prediction exploiting secondary
  structure information.
\newblock In {\em Proc of CILC'04}, pages 16--17, 2004.

\bibitem{Park:95}
Britt~H. Park and Michael Levitt.
\newblock The complexity and accuracy of discrete state models of protein
  structure.
\newblock {\em J Mol Biol}, 249:493--507, 1995.

\bibitem{Pesant:99}
Gilles Pesant and Michel Gendreau.
\newblock A constraint programming framework for local search methods.
\newblock {\em Journal of Heuristics}, 5(3):255--279, 1999.

\bibitem{Regin:94}
Jean-Charles R\'egin.
\newblock A filtering algorithm for constraints of difference in {CSPs}.
\newblock In {\em Proc. of 12th National Conference on AI}, pages 362--367,
  1994.

\bibitem{Sali:94b}
A.~Sali, E.~Shakhnovich, and M.~Karplus.
\newblock Kinetics of protein folding. {A} lattice model study of the
  requirements for folding to the native state.
\newblock {\em J Mol Biol}, 235(5):1614--1636, 1994.

\bibitem{Ullah:09}
Abu~Dayem Ullah, Leonidas Kapsokalivas, Martin Mann, and Kathleen Steinh\"ofel.
\newblock Protein folding simulation by two-stage optimization.
\newblock In {\em Proc. of ISICA'09}, CCIS, Wuhan, China, Oct 2009. Springer.
\newblock (accepted).

\bibitem{Wang:03}
G.~Wang and Roland~L. Dunbrack.
\newblock Pisces: a protein sequence culling server.
\newblock {\em Bioinformatics}, 19(12):1589--91, 2003.

\bibitem{Zhang:07}
J.~Zhang, S.~C. Kou, and J.~S. Liu.
\newblock Biopolymer structure simulation and optimization via fragment
  regrowth {Monte Carlo}.
\newblock {\em J Chem Phys}, 126(22):225101, 2007.

\end{thebibliography}


\end{document}